\documentclass[utf8]{frontiersSCNS} 

\usepackage{url,hyperref,lineno,microtype,subcaption}
\usepackage[onehalfspacing]{setspace}
\usepackage{chemformula}
\usepackage{longtable}
\usepackage{multirow}

\def\keyFont{\fontsize{8}{11}\helveticabold }
\def\firstAuthorLast{Rivilla {et~al.}} 
\def\Authors{V\'ictor M. Rivilla\,$^{1,*}$, 
Izaskun Jim\'enez-Serra\,$^{1}$,
Jes\'us Mart\'in-Pintado\,$^{1}$,
Laura Colzi\,$^{1}$,
Bel\'en Tercero\,$^{2}$,
Pablo de Vicente\,$^{3}$,
Shaoshan Zeng\,$^{4}$,
Sergio Mart\'in\,$^{5,6}$,
Juan Garc\'ia de la Concepci\'on\,$^{1}$,
Luca Bizzocchi\,$^{7}$,
Mattia Melosso\,$^{7,8}$,
Fernando Rico-Villas\,$^{1}$,
Miguel A. Requena-Torres\,$^{9,10}$
}
\def\Address{$^{1}$Centro de Astrobiolog\'{\i}a (CSIC-INTA), Ctra Ajalvir km 4, 28850, Torrej\'{o}n de Ardoz, Madrid, Spain
$^{2}$Observatorio Astronómico Nacional (OAN-IGN), Calle Alfonso XII, 3, 28014 Madrid, Spain
$^{3}$Observatorio de Yebes (OY-IGN), Cerro de la Palera SN, Yebes, Guadalajara, Spain
$^{4}$Star and Planet Formation Laboratory, Cluster for Pioneering Research, RIKEN, 2-1 Hirosawa, Wako, Saitama, 351-0198, Japan
$^{5}$European Southern Observatory, ALMA Department of Science, Alonso de Córdova 3107, Vitacura 763 0355, Santiago, Chile
$^{6}$Joint ALMA Observatory, Department of Science Operations, Alonso de Córdova 3107, Vitacura 763 0355, Santiago, Chile
$^{7}$Department of Chemistry ``Giacomo Ciamician'', University of Bologna, Via F. Selmi 2, Bologna, 40126, Italy
$^{8}$Scuola Superiore Meridionale, Universit\`{a} di Napoli Federico II, Largo San Marcellino 10, Naples, 80138, Italy
$^{9}$University of Maryland, Department of Astronomy, College Park, ND 20742-2421, USA
$^{10}$Department of Physics, Astronomy and Geosciences, Towson University, Towson, MD 21252, USA
}
\def\corrAuthor{Corresponding Author}

\def\corrEmail{vrivilla@cab.inta-csic.es}

\begin{document}
\onecolumn
\firstpage{1}

\title[Molecular precursors of the RNA-world in space: new nitriles towards G+0.693]{Molecular precursors of the RNA-world in space: new nitriles in the G+0.693-0.027 molecular cloud} 

\author[\firstAuthorLast ]{\Authors} 
\address{} 
\correspondance{} 

\extraAuth{}

\maketitle

\begin{abstract}

\section{}

Nitriles play a key role as molecular precursors in prebiotic experiments based on the RNA-world scenario for the origin of life. These chemical compounds could have been partially delivered to the young Earth from extraterrestrial objects, stressing the importance of establishing the reservoir of nitriles in the interstellar medium. 
We report here the detection towards the molecular cloud G+0.693-0.027 of several nitriles, including cyanic acid (HOCN), and three C$_4$H$_3$N isomers (cyanoallene, CH$_2$CCHCN; propargyl cyanide, HCCCH$_2$CN; and cyanopropyne (CH$_3$CCCN), and the tentative detections of cyanoformaldehyde (HCOCN), and glycolonitrile (HOCH$_2$CN). We have also performed the first interstellar search of cyanoacetaldehyde (HCOCH$_2$CN), which was not detected.
Based on the derived molecular abundances of the different nitriles in G+0.693-0.027 and other interstellar sources, 
we have discussed their formation mechanisms in the ISM. 
We propose that the observed HOCN abundance in G+0.693-0.027 is mainly due to surface chemistry and subsequent shock-induced desorption, while HCOCN might be mainly formed through gas-phase chemistry. In the case of HOCH$_2$CN, several grain-surface routes from abundant precursors could produce it. The derived abundances of the three C$_4$H$_3$N isomers in G+0.693-0.027 are very similar, and also similar to those previously reported in the dark cold cloud TMC-1. This suggests that the three isomers are likely formed through gas-phase chemistry from common precursors, possibly unsaturated hydrocarbons (CH$_3$CCH and CH$_2$CCH$_2$) that react with the cyanide radical (CN).
The rich nitrile feedstock found towards G+0.693-0.027 confirms that interstellar chemistry is able to synthesize in space molecular species that could drive the prebiotic chemistry of the RNA-world.

\tiny
 \keyFont{ \section{Keywords:} astrochemistry, Molecules - ISM, Molecular clouds, RNA-world, prebiotic chemistry} 
\end{abstract}

\section{Introduction}

Life on Earth appeared about 3.8 billion years ago, around 700 Myr after the formation of the planet (\citealt{pearce2018}), but we still do not know the mechanisms that made it possible. One of the most supported hypotheses for the origin of life is known as the RNA world (\citealt{gilbert1986}), in which RNA could have performed both metabolic and genetic roles. The process by which inert matter generated first the building blocks of RNA, ribonucleotides, and ultimately RNA itself, remains a mystery.
Recent laboratory experiments mimicking prebiotic conditions have shown that ribonucleotides could be synthesized starting from simple molecules (e.g. \citealt{powner2009,patel2015,becker2019}).
A plausible origin of this prebiotic material is extraterrestrial delivery (\citealt{oro1961,chyba1992,cooper2001}) during the heavy bombardment of meteorites and comets that occurred around 3.9 billions ago (\citealt{marchi2014}). 
These basic molecular precursors may have been already formed prior to the formation of the Solar System, in its parental molecular cloud, through the chemistry that takes place in the interstellar medium (ISM). 
Therefore, the study of the molecular complexity of the ISM can provide us an illustrative view of the chemical reservoir that could have contributed to feed the prebiotic chemistry on the primitive Earth, and could potentially develop similar processes in other places in the Galaxy under favourable Earth-like planetary environments.

In the last decades, and especially in the last years, astrochemistry has shown that interstellar chemistry is able to synthesize building blocks of key biomolecules. Several of the precursors of ribonucleotides spotted by the prebiotic experiments in the laboratory have been detected in the ISM, like cyanoacetylene (HC$_3$N, \citealt{turner1971}), cyanamide (NH$_2$CN, \citealt{turner_microwave_1975}), glycolaldehyde (CH$_2$OHCHO, \citealt{hollis2004}, urea (NH$_2$CONH$_2$, \citealt{belloche2019}), hydroxylamine (NH$_2$OH, \citealt{rivilla2020b}), and 1,2-ethenediol ((CHOH)$_2$; \citealt{rivilla2022a}).
Among the key simple molecular precursors required for the RNA world, numerous works have stressed the dominant role of a particular family of compounds, nitriles, which are molecules with the \ch{C+N} moiety. This simple but highly versatile functional group offers a unique potential to build-up molecular complexity and activate efficiently the formation of ribonucleotides (\citealt{powner2009,powner2010,patel2015,mariani2018,becker2019,menor2020}), and also other key biomolecules such as peptides or nucleobases (\citealt{menor2012,canavelli2019,foden2020}).

With the aim of extending our knowledge on the chemistry of nitriles in the ISM, in this work we have searched for more nitriles towards the molecular cloud G+0.693-0.027 (hereafter G+0.693), including some with increasing complexity that have been proposed as important precursors of prebiotic chemistry. This cloud, located in the Sgr B2 region of the center of our Galaxy, is one of the most chemically rich sources in the ISM. Numerous nitrogen-bearing species, including nitriles, have been detected (see \citealt{zeng2018,rivilla2019b,rivilla2021b}): cyanoacetylene (HC$_3$N), acetonitrile (CH$_3$CN), cyanamide (NH$_2$CN), the cyanomethyl radical (H$_2$CCN), cyanomethanimine (HNCHCN), and the cyanomidyl radical (HNCN).
In this work we report the detection of cyanic acid (HOCN), the tentative detections of 
glycolonitrile (HOCH$_2$CN) and cyanoformaldehyde (HCOCN), and the first interstellar search of cyanoacetaldeyde (HCOCH$_2$CN) in the ISM, for which we provide an abundance upper limit.
We have also searched for three unsaturated carbon-chain nitriles, the C$_4$H$_3$N isomers. We report the detection of cyanopropyne (CH$_3$CCCN), and the second detections in the ISM of cyanoallene (CH$_2$CCHCN) and propargyl cyanide (HCCCH$_2$CN), detected previously only towards the TMC-1 dark cloud (\citealt{lovas2006_cyanoallene,mcguire2020early,marcelino2021}).
In Section \ref{sec:observations} we present the data of the observational survey, in Section \ref{sec:results} we describe the line identification and analysis, and present the results of the line fitting, and in Section  \ref{sec:discussion} we discuss about the interstellar chemistry of the different species and their possible roles in prebiotic chemistry.

\section{Observations}
\label{sec:observations}

A high sensitivity spectral survey was carried out towards G+0.693. We used both IRAM 30m telescope (Granada, Spain) and Yebes 40m telescope (Guadalajara, Spain). The observations were centred at $\alpha$(J2000.0)$\,$=$\,$17$^h$47$^m$22$^s$, and $\delta$(J2000.0)$\,$=$\,-\,$28$^{\circ}$21$'$27$''$. The position switching mode was used in all the observations with the off position located at $\Delta\alpha$~=~$-885$'', $\Delta\delta$~=~$290$'' from the source position.
During the IRAM 30m observations the dual polarisation receiver EMIR was connected to the fast Fourier transform spectrometers (FFTS), which provided a channel width of 200 kHz. In this work we have used data covering the spectral windows from 71.8 to 116.7 GHz, 124.8 to 175.5 GHz, and 199.8$-$238.3 GHz. The spectra were smoothed to velocity resolutions of 1.0$-$2.6 km s$^{-1}$, depending on the frequency. 
The observations with the Yebes 40m radiotelescope used the Nanocosmos Q-band (7$\,$mm) HEMT receiver (\citealt{tercero2021}). 
The receiver was connected to 16 FFTS providing a channel width of 38 kHz and a bandwidth of 18.5 GHz per polarisation, covering the frequency range between 31.3 GHz and 50.6 GHz. 
The spectra were smoothed to a resolution of 251 kHz, corresponding to velocity resolutions of 1.5$-$2.4 km s$^{-1}$. The noise of the spectra depends on the frequency range, with values in antenna temperature ($T_{A}^{*}$) as low as 1.0 mK, while in some intervals it increases up to 4.0$-$5.0 mK, for the Yebes data, and 1.3 to 2.8 mK (71$-$90 GHz), 1.5 to 5.8 mK (90$-$115 GHz), $\sim$10 mK (115$-$116 GHz), 3.1 to 6.8 mK (124$-$175 GHz), and 4.5 to 10.6 mK (199$-$238 GHz), for the IRAM 30m data. 
The line intensity of the spectra was measured in units of $T_{\mathrm{A}}^{\ast}$ as the molecular emission toward G+0.693 is extended over the beam (\citealt{requena-torres_organic_2006,requena-torres_largest_2008,zeng2018,zeng2020}).

\section{Analysis and results}
\label{sec:results}

\begin{figure}
\begin{center}
\hspace{-14mm}
\includegraphics[width=19cm]{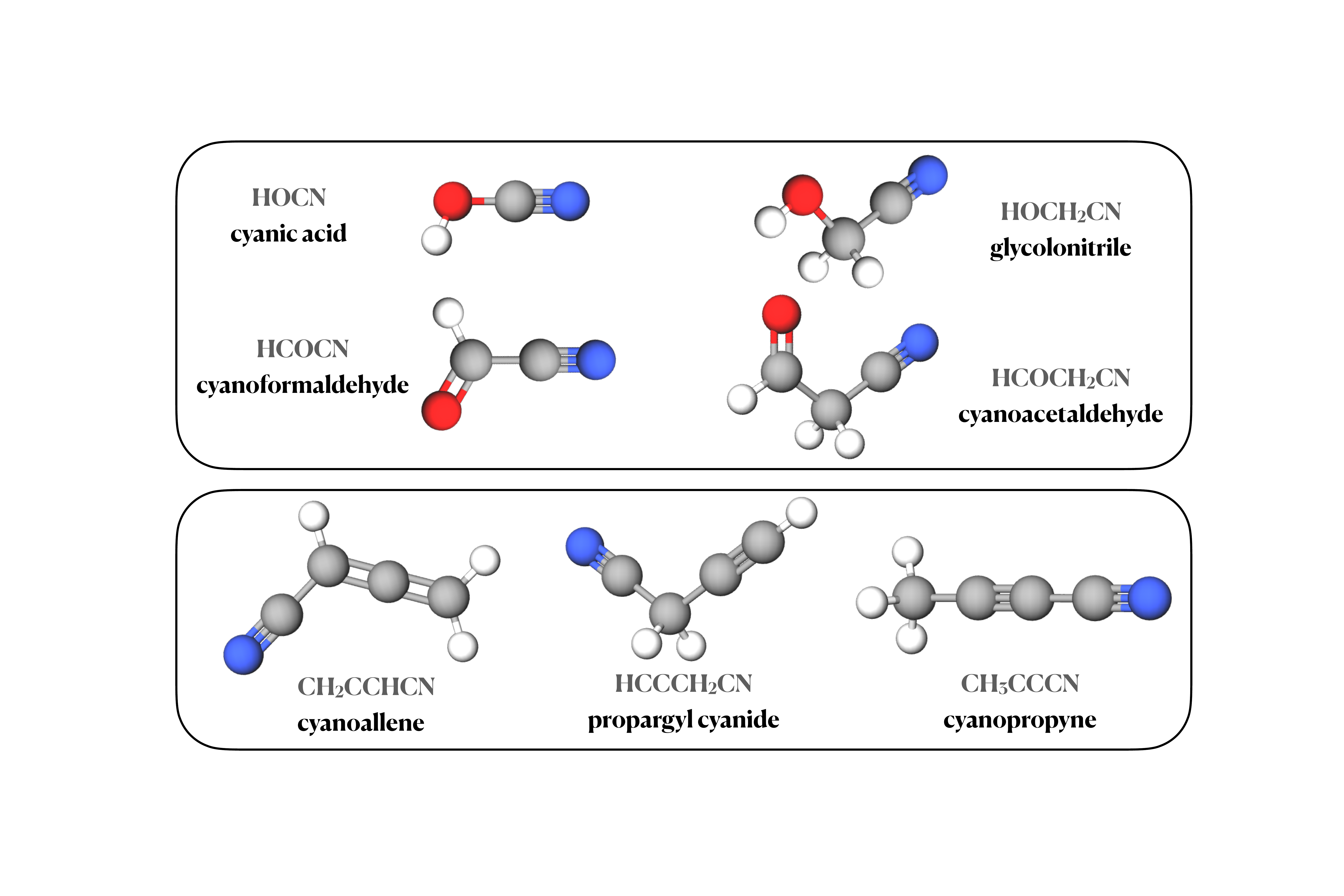}
\end{center}
\vskip-20mm
\caption{Three-dimensional representation of the oxygen-bearing nitriles (upper panel) and the three C$_4$H$_3$N isomers (lower panel) analysed in this work. White, gray, red, and blue corresponds to hydrogen, carbon, oxygen and nitrogen atoms, respectively.}
\label{fig:molecules}
\end{figure}

Figure \ref{fig:molecules} shows the nitriles analysed in this work, which include four oxygen-bearing nitriles: cyanic acid (HOCN), cyanoformaldehyde (or formyl cyanide, HCOCN), glycolonitrile (or 2-hydroxyacetonitrile, HCOCH$_2$CN), cyanoacetaldehyde (or 3-oxopropanenitrile, HCOCH$_2$CN); and three C$_4$H$_3$N isomers: cyanoallene (or 2,3-butadienenitrile, CH$_2$CCHCN), propargyl cyanide (or 3-butynenitrile, HCCCH$_2$CN), and cyanopropyne (or 2-butynenitrile, CH$_3$CCCN). The identification and fitting of the molecular lines were performed using the SLIM (Spectral Line Identification and Modeling) tool within the MADCUBA package{\footnote{Madrid Data Cube Analysis on ImageJ is a software developed at the Center of Astrobiology (CAB) in Madrid; https://cab.inta-csic.es/madcuba/}} (version 09/11/2021; \citealt{martin2019}). SLIM generates synthetic spectra under the assumption of Local Thermodynamic Equilibrium (LTE), using the spectroscopy provided by laboratory experiments assisted by theoretical calculations.  Table \ref{tab:spectroscopy} lists the spectroscopic references of all the molecules analysed. We have used entries from the Cologne Database for Molecular Spectroscopy (CDMS, \citealt{endres2016}), which are based on the laboratory works and theoretical calculations indicated in Table \ref{tab:spectroscopy}. Moreover, we implemented into MADCUBA the spectroscopy of HCOCH$_2$CN from \citet{Mollendal2012}.

\begin{table}
\centering
\tabcolsep 3pt
\caption{Spectroscopy of the molecules analysed in this work. The molecular catalog, number and date of the entry, and the references for the line lists and dipole moments are listed.}
\begin{tabular}{l c c c c c }
\hline
Molecule & Catalog  &   Entry & Date & Line list ref. & Dipole moment ref. \\ 
\hline
HOCN & CDMS & 43510 & May 2009  & \citet{brunken2009} & \citet{brunken2009} \\
HCOCN & CDMS & 55501 & June 2006  & \citet{bogey1995} &\citealt{csaszar1989} \\
HOCH$_2$CN & CDMS & 57512 & March 2017 & \citet{margules_submillimeter_2017} &  \citet{margules_submillimeter_2017}\\
HCOCH$_2$CN& MADCUBA & -- & January 2022 & \citet{Mollendal2012} & \citet{Mollendal2012}\\
\hline
HCCCH$_2$CN & CDMS & 65514 & September 2020 &\citet{demaison1985}  &\citet{demaison1985} \\
&  &  &  & \citet{mcguire2020early}  & \\
CH$_2$CCHCN & CDMS & 65506 &  March 2006 &  \citet{bouchy1973} & \citet{bouchy1973}\\
 &  &  &   &  \citet{schwahn1986} & \\
CH$_3$CCCN & CDMS & 65503 & April 2004 & \citet{moises1982}, & \\
 &  &  &  & \citet{bester1983} & \citet{bester1984}\\
\hline
\end{tabular}
\label{tab:spectroscopy}
\end{table}

To evaluate if the molecular transitions of interest are blended with emission from other species, we have also considered the LTE model that includes the total contribution of all the species that have been identified so far towards G+0.693 (e.g., \citealt{requena-torres_largest_2008,zeng2018,rivilla2019a,rivilla2020b,jimenez-serra2020,rivilla2021a,rivilla2021b,zeng2021,rodriguez-almeida2021a,rodriguez-almeida2021_b,rivilla2022a,rivilla2022b}). To derive the physical parameters of the molecular emission, we used the AUTOFIT tool of SLIM, which finds the best agreement between the observed spectra and the predicted LTE model, and provides the best solution for the parameters, and their associated uncertainties (see details of the formalism used in \citealt{martin2019}). The free parameters of the model are: molecular column density ($N$), excitation temperature ($T_{\rm ex}$), linewidth (or full width at half maximum, FWHM), and velocity ($v_{\rm LSR}$). We have left these four parameters free whenever possible, providing their associated uncertainties. For the cases in which the algorithm used by AUTOFIT does not converge, we have fixed some of them to allow the algorithm to converge. 
In the following, we present the analysis of the different molecules studied. 
For each species, we have applied AUTOFIT using unblended transitions and transitions that, while partially blended with other species already identified in G+0.693, properly reproduces the observed spectra. We note that for all molecules the transitions that are not shown are consistent with the observed spectra, but they are heavily blended with lines from other molecular species or they are too weak to be detected, according to the line intensities predicted by the LTE model.

\begin{table}
\centering
\tabcolsep 6pt
\caption{List of detected transitions of the oxygen-bearing nitriles analysed in this work. We indicate the frequency, quantum numbers, logarithm of the Einstein coefficient ($A_{\rm ul}$), energy of the upper levels of each transition ( E$_{\rm u}$), and information about the possible blending by other identified or unidentified (U) species towards G+0.693.}
\vspace{2mm}
\begin{tabular}{l c c c  c l}
\hline
Molecule & Frequency & Transition & log$A_{\rm ul}$   & E$_{\rm u}$ &   Blending \\
& (GHz) &  $J_{K_a,K_c}$ & (s$^{-1}$) & (K) &   \\
\hline
HOCN  &  41.9508371 &   2$_{0,2}-$1$_{0,1}$   &  -5.3239  &  3.0   & unblended  \\  
HOCN  &  83.5383960 &   4$_{1,4}-$3$_{1,3}$   &  -4.4087 &  42.2    &  blended with U \\ 
HOCN  &  83.9005702$^{a}$ &   4$_{0,4}-$3$_{0,3}$    & 	 -4.3750  &  10.1 & unblended  \\ 
HOCN  &  84.2524547 &   4$_{1,3}-$3$_{1,2}$   &  -4.3976 &   42.3   & blended with HCCCH$_2$CN  \\ 
HOCN  & 104.8746777$^{a}$ &   5$_{0,5}-$4$_{0,4}$   &  -4.0746    &  15.1 &  unblended \\ 
HOCN  & 125.8480951 &   6$_{0,6}-$5$_{0,5}$   & -3.8304 &   21.1   &  unblended \\ 
HOCN  & 146.8206846$^{a}$ &   7$_{0,7}-$6$_{0,6}$  & -3.6248   &  28.1  &  unblended \\   
HOCN  & 167.7923140$^{b}$ &   8$_{0,8}-$7$_{0,7}$   & -3.4472   &  36.2    &  unblended \\ 
\hline
HCOCN      & 72.1192555   &  1$_{1,1}-$0$_{0,0}$   &  -5.0767   & 3.5  & blended \\ 
HCOCN      & 81.433113  &  2$_{1,2}-$1$_{0,1}$    &  -4.9642    &  4.4 & blended  with NH$_2$CH$_2$CH$_2$OH \\
HCOCN      & 90.5710141  &  3$_{1,3}-$2$_{0,2}$    &  -4.8461    & 5.7  &   unblended \\ 
HCOCN      & 99.5348108   &  4$_{1,4}-$3$_{0,3}$   & -4.7342     &  7.6 & unblended  \\ 
HCOCN      & 108.3274964 &  5$_{1,5}-$4$_{0,4}$    & -4.6301     &  9.8 &  part. blended with HCCO and U  \\ 
HCOCN      & 207.3132961   &  2$_{2,0}-$1$_{1,1}$    &  -3.7478    & 13.4  &  blended with U  \\
\hline
HOCH$_2$CN   & 35.934379   &  4$_{1,4}-$3$_{1,3}$   & -5.8654   &   5.7  & unblended  \\ 
HOCH$_2$CN   & 37.781601   & 4$_{1,3}-$3$_{1,2}$   & -5.8001   &   5.9   & blended with c-C$_2$H$_4$O and U  \\ 
HOCH$_2$CN   & 44.907438    &  5$_{1,5}-$4$_{1,4}$  & -5.5549  &  7.9    &  blended with $t-$HCOOH \\ 
HOCH$_2$CN   & 45.975528   &  5$_{0,5}-$4$_{0,4}$   & -5.5069   & 6.6  & unblended   \\
HOCH$_2$CN   & 75.463333   & 8$_{1,7}-$7$_{1,6}$    & -4.8528     & 17.7  & blended with $s$-C$_2$H$_5$CHO  \\ 
\hline
\end{tabular}
\label{tab:transitions_O}
{\small
{\\ (a) Transition detected in \citet{zeng2018}.}
{\\ (b) Transition tentatively detected in \citet{zeng2018}.}
}
\end{table}

\begin{figure}
\begin{center}
\includegraphics[width=18cm]{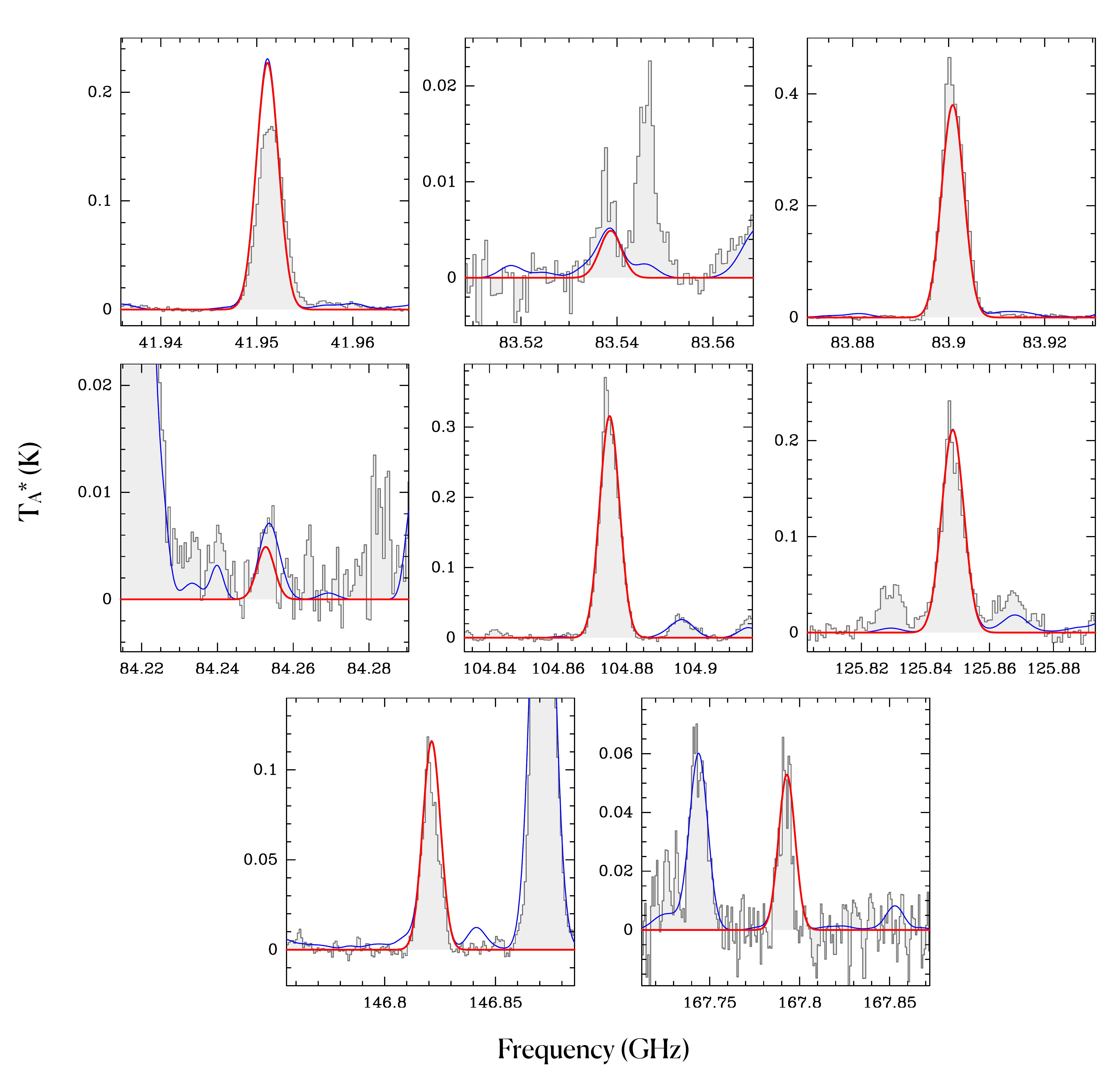}
\vspace{-8mm}
\end{center}
\caption{Selected cyanic acid (HOCN) transitions (see Table \ref{tab:transitions_O}) detected towards the G+0.693 molecular cloud. The best LTE fit derived with MADCUBA for the HOCN emission is shown with a red curve, while the blue curve shows the total emission considering all the species identified towards this molecular cloud. The y-axis shows the line intensity in antenna temperature scale ($T_{\rm A}^*$) in Kelvin, and the x-axis shows the frequency in GHz.}
\label{fig:molecules_hocn}
\end{figure}

\begin{figure}
\begin{center}
\includegraphics[width=18cm]{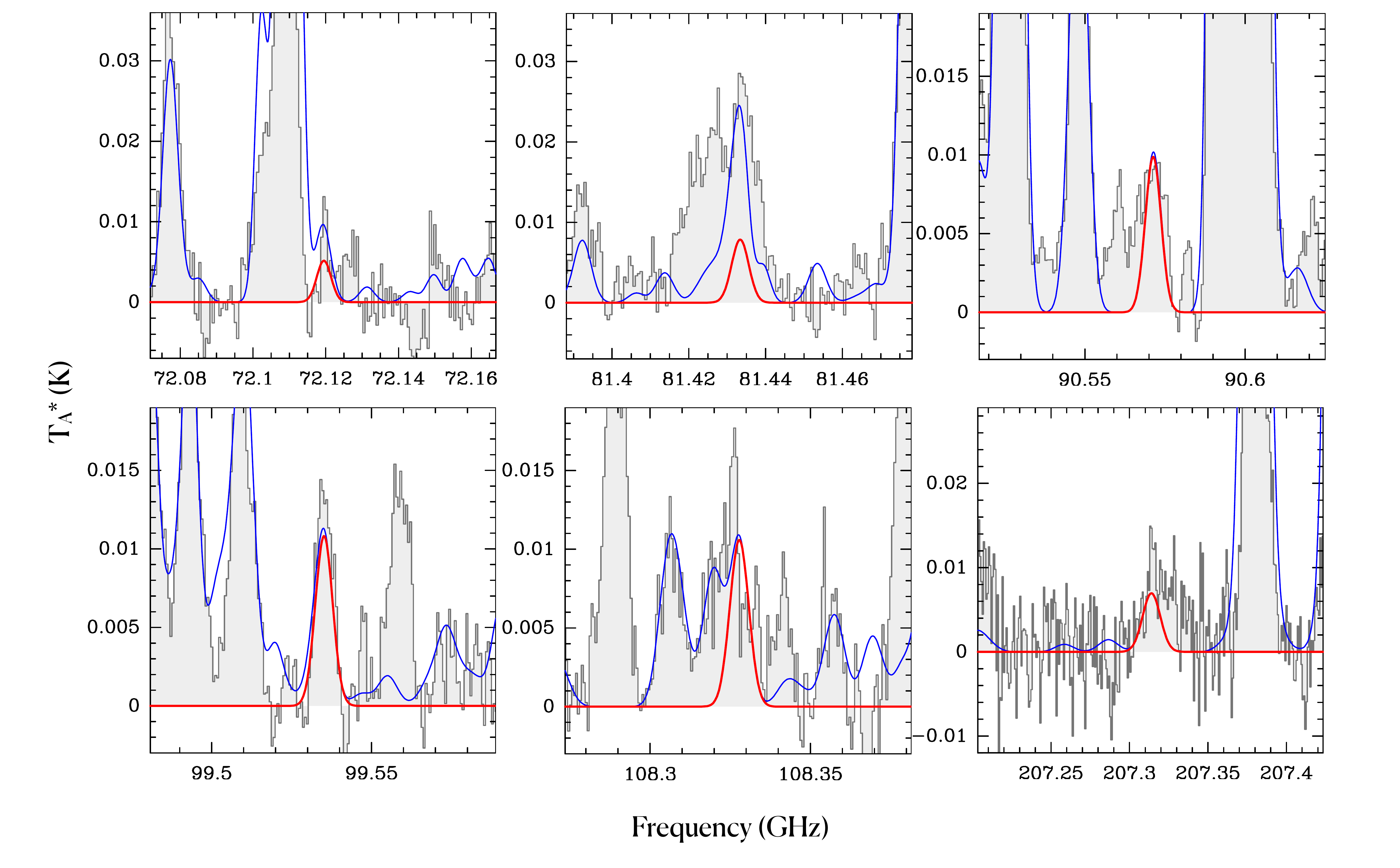}
\end{center}
\vspace{-5mm}
\caption{Selected cyanoformaldehyde (HCOCN) transitions (see Table \ref{tab:transitions_O}) detected towards the G+0.693 molecular cloud. The best LTE fit derived with MADCUBA for the HCOCN emission is shown with a red curve, while the blue curve shows the total emission considering all the species identified towards this molecular cloud. The y-axis shows the line intensity in antenna temperature scale ($T_{\rm A}^*$) in Kelvin, and the x-axis shows the frequency in GHz.}
\label{fig:molecules_hcocn}
\end{figure}

\begin{figure}
\begin{center}
\includegraphics[width=18cm]{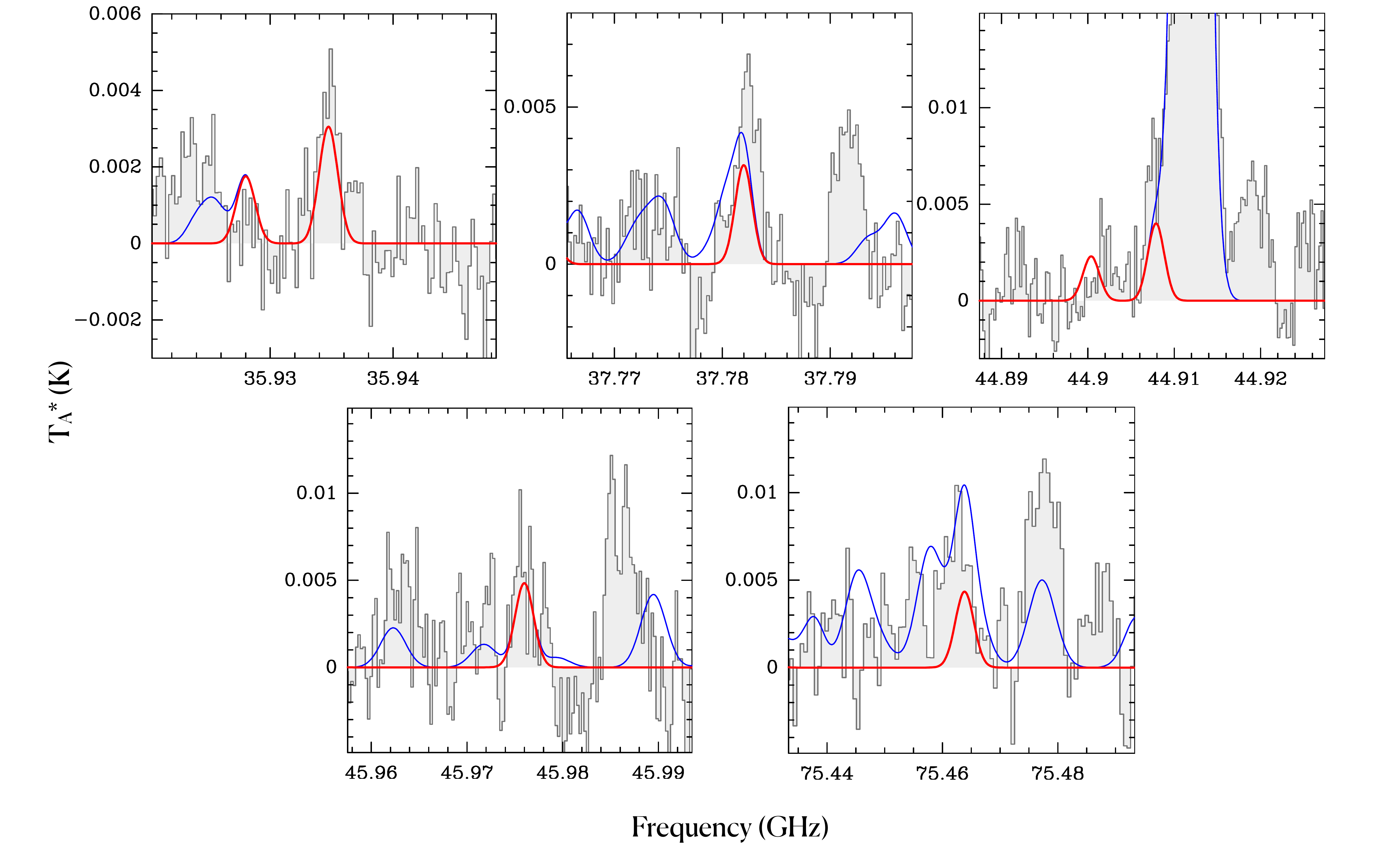}
\end{center}
\vspace{-5mm}
\caption{Selected transitions of glycolonitrile (HOCH$_2$CN; see Table \ref{tab:transitions_O}) detected towards the G+0.693 molecular cloud. The best LTE fit derived with MADCUBA for the HOCH$_2$CN emission is shown with a red curve, while the blue curve shows the total emission considering all the species identified towards this molecular cloud. The y-axis shows the line intensity in antenna temperature scale ($T_{\rm A}^*$) in Kelvin, and the x-axis shows the frequency in GHz.} 
\label{fig:molecules_hoch2cn}
\end{figure}
%
\vspace{6mm}

\subsection{Oxygen-bearing nitriles}

\vspace{2mm}
\subsubsection{Cyanic acid (HOCN) and cyanoformaldehyde (HCOCN)}

HOCN was already reported towards G+0.693 by \citet{brunken_interstellar_2010} (their source Sgr B2 (20,100)\footnote{The position of this source is offset in ($\alpha$, $\delta$) by (20$''$,100$''$) with respect to that of Sgr B2(M), see \citet{brunken_interstellar_2010}.}), and also by \citet{zeng2018} using in both cases less sensitive observations. We provide here a new analysis using deeper observations. 
We have detected six transitions of this species that are completely unblended, which are shown in Figure \ref{fig:molecules_hocn}, and listed in Table \ref{tab:transitions_O}. These transitions include the three transitions identified by \citet{zeng2018}, the confirmation of the 8$_{0,8}-$7$_{0,7}$ transition tentatively detected in that work (see their Figure B15), and two new transitions (Table \ref{tab:transitions_O}). 
The best LTE fit derived by MADCUBA, where all parameters were left free, is shown in Figure \ref{fig:molecules_hocn}, and the derived physical parameters are presented in Table \ref{tab:parameters}.
We obtained a column density of (2.13$\pm$0.04)$\times$10$^{13}$ cm$^{-2}$ (Table \ref{tab:parameters}), which translates into a molecular abundance with respect to molecular hydrogen of 1.6$\times$10$^{-10}$, using $N_{\rm H_2}$=1.35$\times$10$^{23}$ cm$^{-2}$ from \citet{martin_tracing_2008}. The results are consistent, within the uncertainties, with those derived by \citet{zeng2018}.

We also report here the first tentative detection of HCOCN towards G+0.693. 
Figure \ref{fig:molecules_hcocn} shows that the 3$_{1,3}-$2$_{0,2}$ (90.5710141 GHz) and 4$_{1,4}-$3$_{0,3}$ (99.5348108 GHz) transitions are unblended, while other transitions are partially blended with other species (Table \ref{tab:transitions_O}). 
To perform the fit, we fixed $T_{\rm ex}$, FWHM,  and $v_{\rm LSR}$ to the ones derived from HOCN. We obtained a column density of (0.76$\pm$0.11)$\times$10$^{13}$ cm$^{-2}$, almost one order of magnitude lower than the upper limit reported by \citet{zeng2018} of $<$6$\times$10$^{13}$ cm$^{-2}$ towards G+0.693. The derived molecular abundance is 6$\times$10$^{-11}$, which is very similar to that found in the TMC-1 dark cloud by \citet{cernicharo2021_sulfur}. The HOCN/HCOCN ratio is $\sim$2.8.
\vspace{8mm}

\subsubsection{Glycolonitrile (HOCH$_2$CN)}

This species is also tentatively detected towards G+0.693.
We show in Figure \ref{fig:molecules_hoch2cn} two molecular transitions of HOCH$_2$CN that are unblended (Table \ref{tab:transitions_O}), and those partially blended with other species already identified in this cloud. 
To perform the fit, we fixed $T_{\rm ex}$ and FWHM to the ones derived for HOCN, and used $v_{\rm LSR}$=67 km s$^{-1}$, which best reproduces the velocity of the two unblended transitions. 
We obtained a column density of (0.8$\pm$0.2)$\times$10$^{13}$ cm$^{-2}$ (Table \ref{tab:parameters}), and a molecular abundance of 6$\times$10$^{-11}$, very similar to that of HOCN.

\vspace{8mm}
\subsubsection{Cyanoacetaldehyde (HCOCH$_2$CN)}

This molecule is not currently included in any of the commonly used molecular databases such as CDMS or the Jet Propulsion Laboratory catalog (JPL; \citealt{pickett1998}).
The conformational energy landscape of HCOCH$_2$CN and the effects of the large amplitude motions on its rotational spectrum have been described in detail by \citet{Mollendal2012}. We have used the spectroscopic information provided in this work to implement it into MADCUBA.
The most stable rotamer (referred to as species~I in the cited reference) possesses two equivalent positions in the electronic energy potential function for rotation about its 
C$_1$--C$_2$ bond (see figure 1 of \citealt{Mollendal2012}).
They are separated by a barrier of 0.84\,kJ\,mol$^{-1}$ (computed at MP2 level) at the exact antiperiplanar conformation.
Large amplitude vibrations and tunneling for the torsion about the C$_1$--C$_2$ bond leads 
to the existence of two closely spaced energy levels for the ground state labelled with a 
plus sign ($+$) for the lowest-energy level and with a minus sign ($-$) for the 
higher-energy level. These two states are separated by an energy difference $\Delta E/h$ of $\sim$58.8 GHz.
For the present spectral calculation we have reanalysed the rotational transitions reported 
by \citet{Mollendal2012} using the same set of spectroscopic parameters employed in their
fit~1 (see their Table~4).
The rest-frequencies have then been computed in the $J$=0--70 interval with $K_{a\, max} = 50$.
Theoretical values of dipole moments $\mu_a = 0.932$\,D, $\mu_b = 1.574$\,D, and $\mu_c = 1.274$\,D, computed at CCSD level \citep{Mollendal2012} have been employed.
All the calculations have been performed with the CALPGM suite of programs \cite{Pickett1991}.

This species is not detected towards G+0.693. We have derived an upper limit for its abundance using the brightest transition according to the LTE model that are unblended, namely the 6$_{2,5}$-5$_{1,4}$ transition at 101.598576 GHz. 
MADCUBA calculates the upper limit of the column density using the 3$\sigma$ value of the integrated intensity (see details in \citealt{martin2019}).
We have used the same $T_{\rm ex}$, FWHM, and $v_{\rm LSR}$ used for HOCH$_2$CN. We obtained an upper limit of the HCOCH$_2$CN abundance of $<$2.7$\times$10$^{-10}$ (Table \ref{tab:parameters}).

\vspace{5mm}

\subsection{C$_4$H$_3$N isomers}

We report in this section the first detection towards G+0.693 of cyanoallene (CH$_2$CCHCN), propargyl cyanide (HCCCH$_2$CN), and cyanopropyne (CH$_3$CCCN).

\begin{figure}
\begin{center}
\includegraphics[width=18cm]{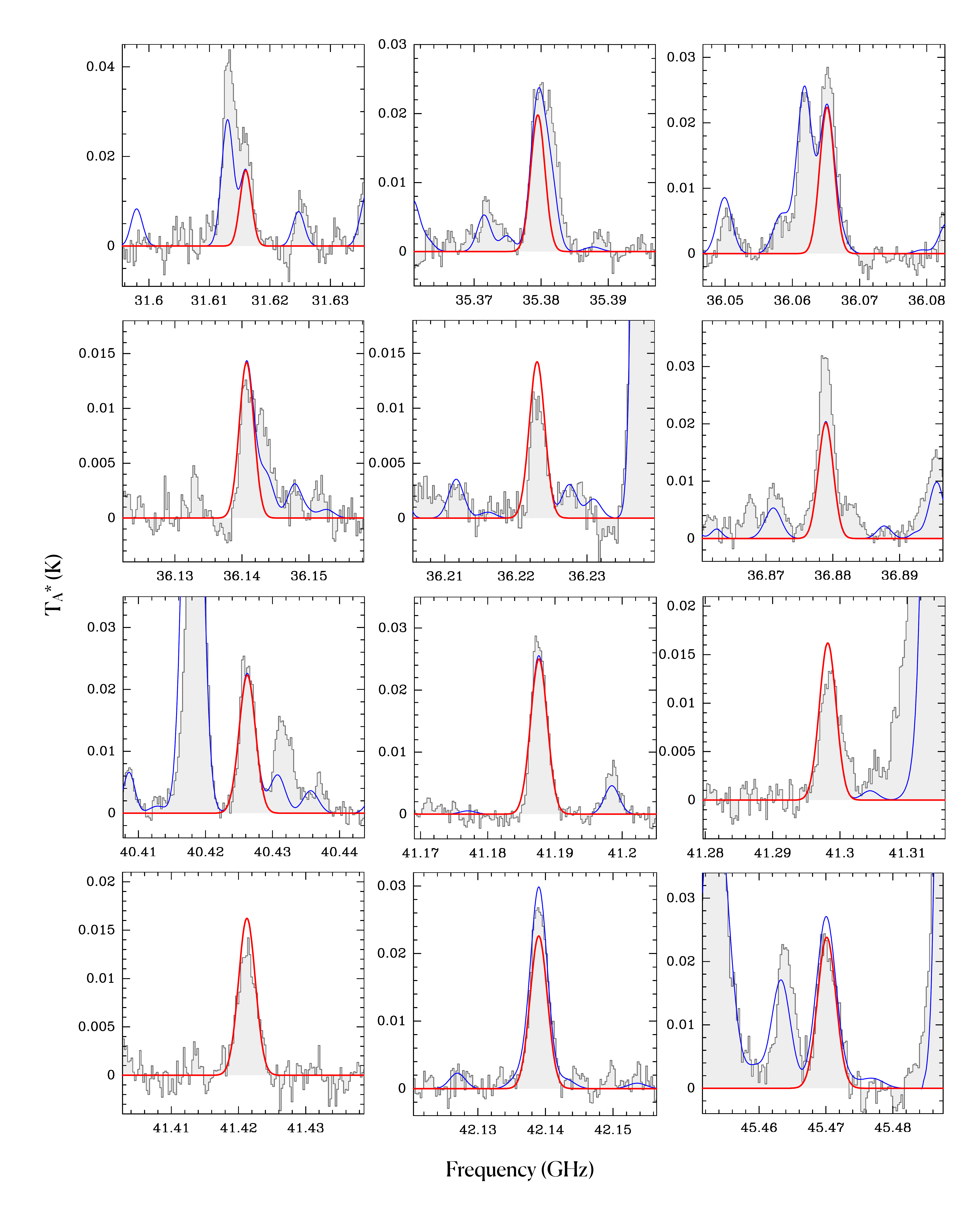}
\end{center}
\vspace{-7mm}
\caption{Selected transitions of cyanoallene (CH$_2$CCHCN; see Table \ref{tab:transitions_isomers}) detected towards the G+0.693 molecular cloud. The best LTE fit derived with MADCUBA for the CH$_2$CCHCN emission is shown with a red curve, while the blue curve shows the total emission considering all the species identified towards this molecular cloud. The y-axis shows the line intensity in antenna temperature scale ($T_{\rm A}^*$) in Kelvin, and the x-axis shows the frequency in GHz.}
\label{fig:molecules_ch2cchcn}
\end{figure}

\begin{figure}
\addtocounter{figure}{-1}
\begin{center}
\includegraphics[width=18cm]{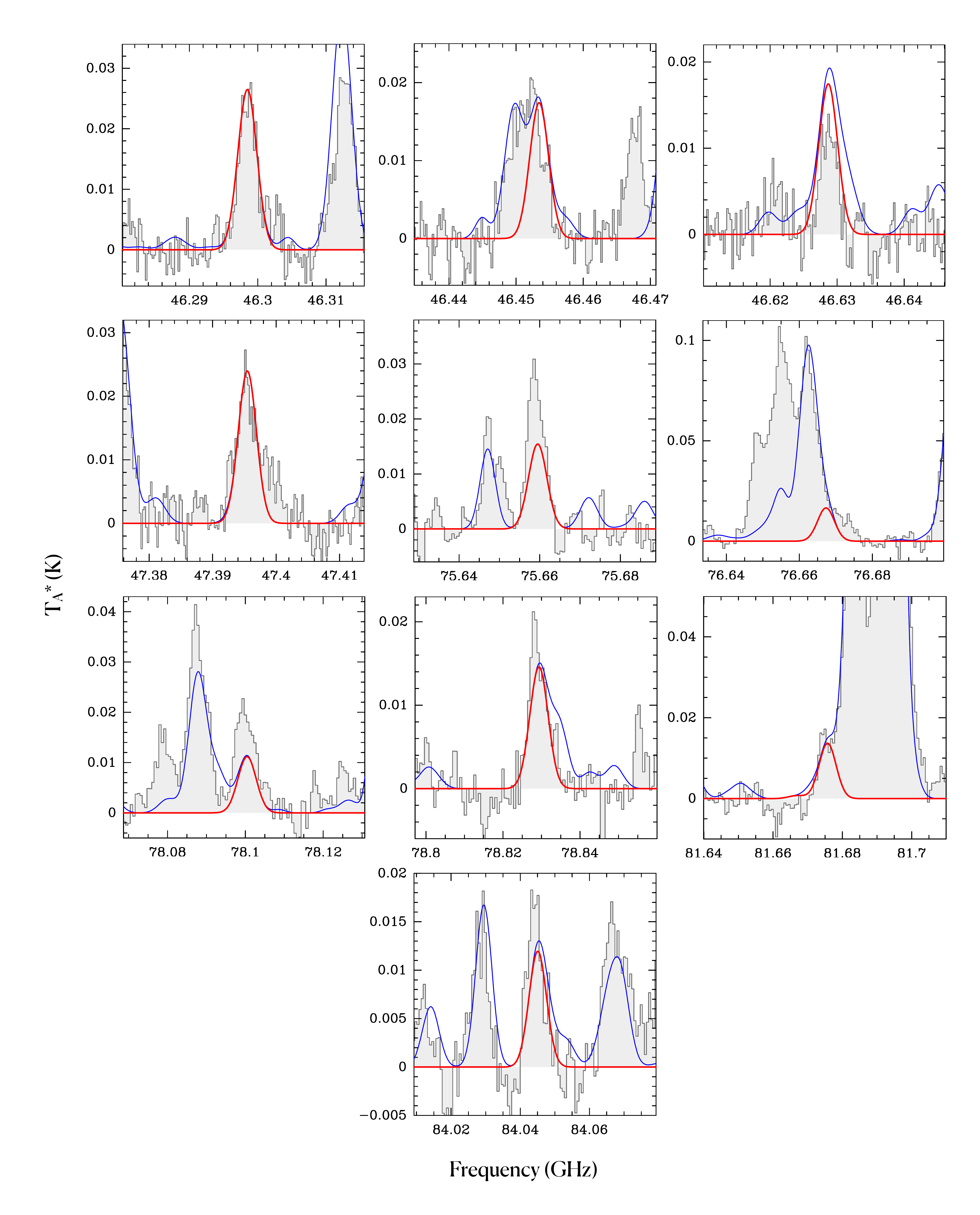}
\end{center}
\caption{Continued.}
\label{fig:molecules_ch2cchcn}
\end{figure}

\begin{figure}
\begin{center}
\includegraphics[width=18cm]{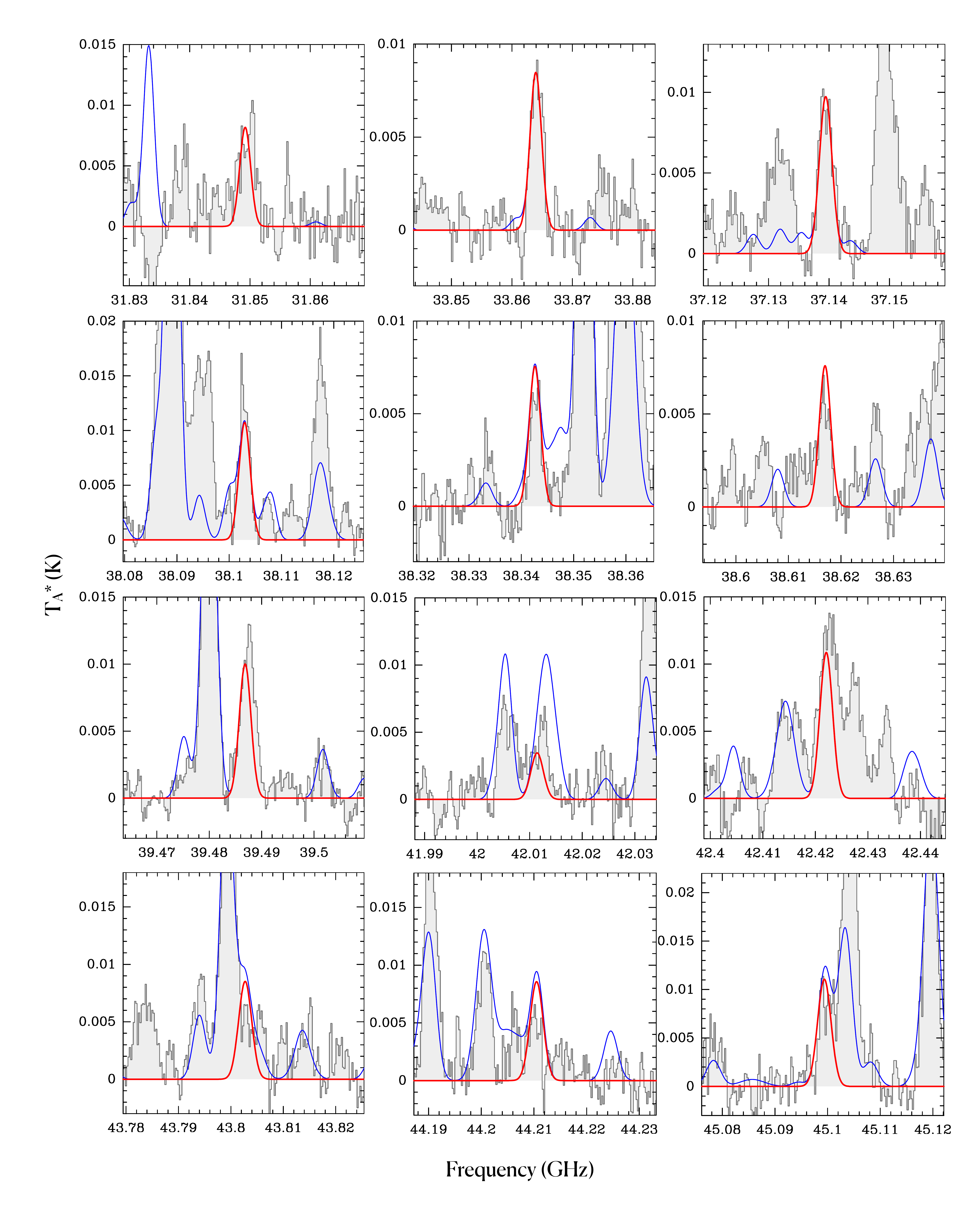}
\end{center}
\vspace{-7mm}
\caption{Selected transitions of propargyl cyanide (HCCCH$_2$CN; see Table \ref{tab:transitions_isomers}) detected towards the G+0.693 molecular cloud. The best LTE fit derived with MADCUBA for the HCCCH$_2$CN emission is shown with a red curve, while the blue curve shows the total emission considering all the species identified towards this molecular cloud. The y-axis shows the line intensity in antenna temperature scale ($T_{\rm A}^*$) in Kelvin, and the x-axis shows the frequency in GHz.}
\label{fig:molecules_pgcn}
\end{figure}

\begin{figure}
\begin{center}
\includegraphics[width=18cm]{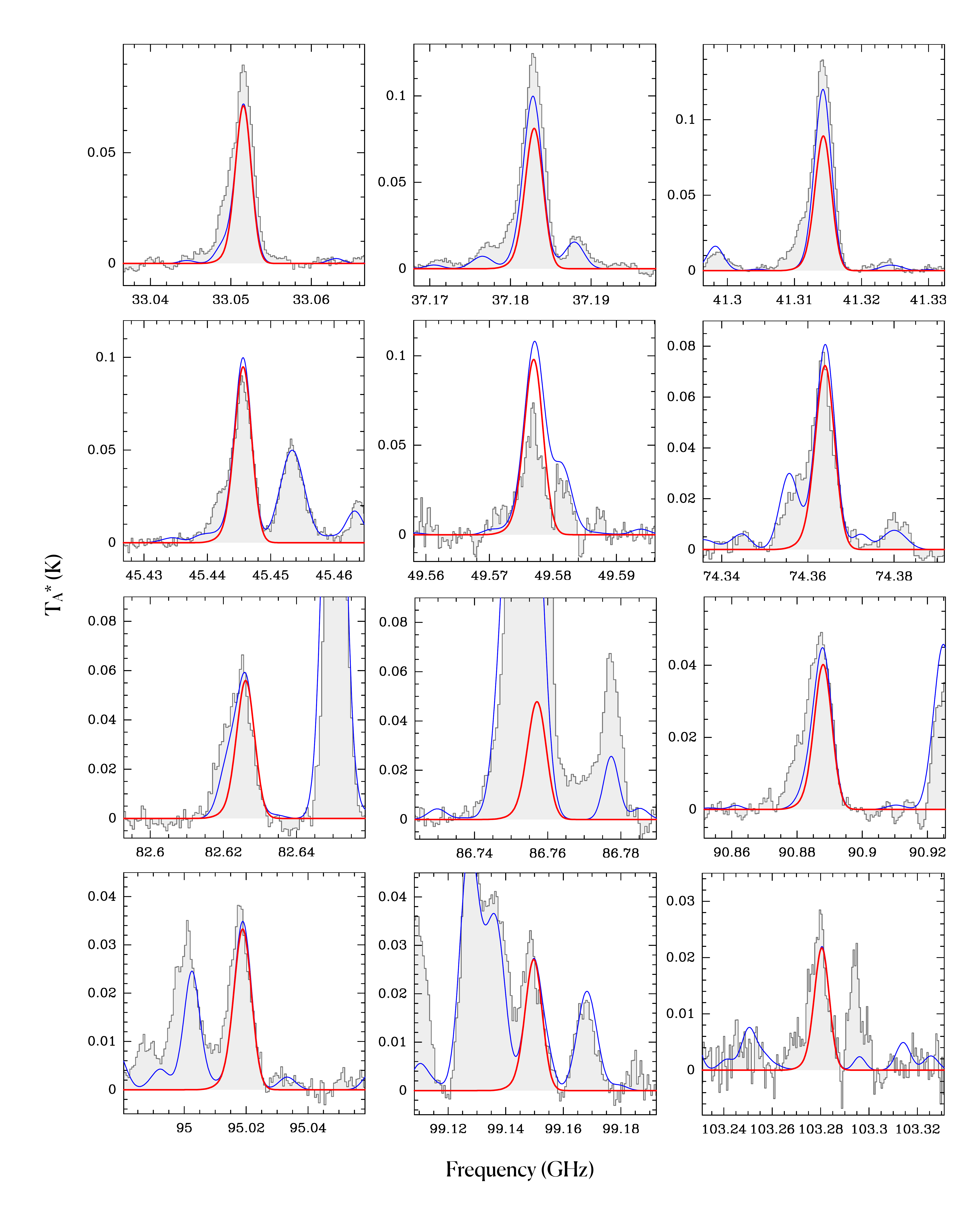}
\end{center}
\vspace{-7mm}
\caption{Selected transitions of cyanopropyne (CH$_3$CCCN; see Table \ref{tab:transitions_isomers}) detected towards the G+0.693 molecular cloud. The best LTE fit derived with MADCUBA for the CH$_3$CCCN emission is shown with a red curve, while the blue curve shows the total emission considering all the species identified towards this molecular cloud. The y-axis shows the line intensity in antenna temperature scale ($T_{\rm A}^*$) in Kelvin, and the x-axis shows the frequency in GHz.}
\label{fig:molecules_ch3c3n}
\end{figure}

\begin{figure}
\addtocounter{figure}{-1}
\begin{center}
\includegraphics[width=7.2cm]{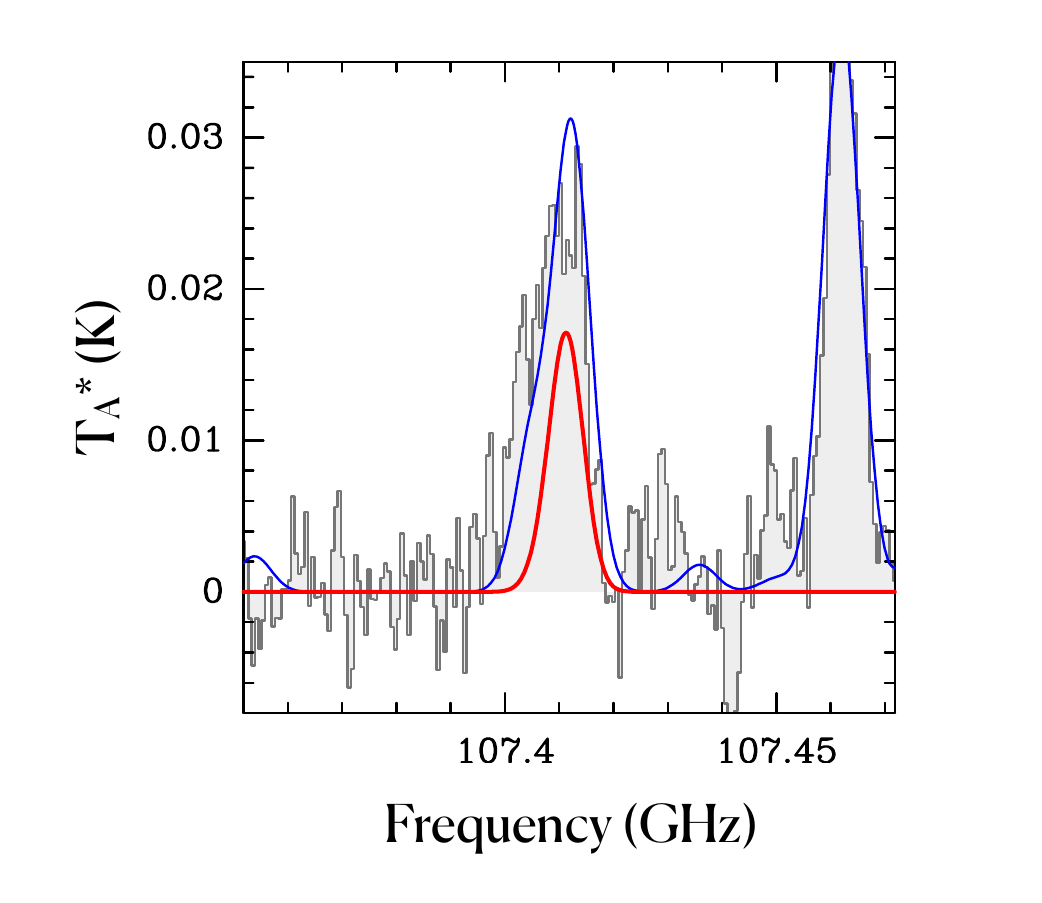}
\end{center}
\vspace{-7mm}
\caption{Continued.}
\label{fig:molecules_ch3c3n}
\end{figure}

\vspace{3mm}
\subsubsection{Cyanoallene (CH$_2$CCHCN)}

Figure \ref{fig:molecules_ch2cchcn} shows the molecular transitions of CH$_2$CCHCN that are unblended, or only slightly blended with other species already identified in this source, whose spectroscopic information is presented in Table \ref{tab:transitions_isomers}.
The G+0.693 cloud is the second interstellar source where CH$_2$CCHCN has been detected, after the cold cloud TMC-1 (\citealt{lovas2006_cyanoallene,marcelino2021}).
We left $N$, $T_{\rm ex}$, FWHM, and $v_{\rm LSR}$ as free parameters, and obtained a column density of (2.34$\pm$0.06)$\times$10$^{13}$ cm$^{-2}$, and a molecular abundance of 1.7$\times$10$^{-10}$ (Table \ref{tab:parameters}).

\vspace{3mm}
\subsubsection{Propargyl cyanide (HCCCH$_2$CN)}

Figure \ref{fig:molecules_pgcn} shows the molecular transitions of HCCCH$_2$CN that are unblended, or only slightly blended with other species already identified in this cloud, whose spectroscopic information is presented in Table \ref{tab:transitions_isomers}.
As in the case of its isomer CH$_2$CCHCN, G+0.693 is the second interstellar source where HCCCH$_2$CN has been detected, after the cold cloud TMC-1 (\citealt{mcguire2020early,marcelino2021}).
We fixed $T_{\rm ex}$ and FWHM to the values obtained for CH$_2$CCHCN, and left $N$ and $v_{\rm LSR}$ free.
We obtained a column density of (1.77$\pm$0.08)$\times$10$^{13}$ cm$^{-2}$, and a molecular abundance of 1.3$\times$10$^{-10}$. The CH$_2$CCHCN/HCCCH$_2$CN ratio is $\sim$ 1.3.

\begin{longtable}{l c c  c c l}
\caption{. List of observed transitions of the C$_4$H$_3$N isomers analysed in this work. We indicate the frequency, quantum numbers, Einstein coefficient ($A_{\rm ul}$), energy of the upper levels of each transition (E$_{\rm u}$), and information about the possible blending by other identified or unidentified (U) species towards G+0.693.}
\tabularnewline \hline 
Molecule & Frequency & Transition$^{a}$  &  log$A_{\rm ul}$   & E$_{\rm u}$ &   Blending \\
& (GHz) &  & (s$^{-1 }$) & (K) &   \\
\hline
\endfirsthead
\caption{. Continued.}\\ 
\hline 
Molecule & Frequency & Transition$^{a}$   & log$A_{\rm ul}$ & E$_{\rm u}$ &   Blending \\
& (GHz) &  & (s$^{-1 }$) & (K) &   \\
\hline
\endhead
\hline
CH$_2$CCHCN & 31.6156000	     & 6$_{1,5}-$5$_{1,4}$   & -5.5633  &   6.4 & blended with $aGg'$-(CH$_2$OH)$_2$ \\
CH$_2$CCHCN & 35.3790500       & 7$_{1,7}-$6$_{1,6}$   & -5.4086   & 7.9  &  blended with CH$_3$COCH$_3$ \\
CH$_2$CCHCN & 36.0646889     & 7$_{0,7}-$6$_{0,6}$   & -5.3748   & 6.9  &  unblended \\
CH$_2$CCHCN & 36.1402200       & 7$_{2,6}-$6$_{2,5}$   & -5.4089   & 11.4  & unblended  \\ 
CH$_2$CCHCN & 36.2225000        & 7$_{2,5}-$6$_{2,4}$  & -5.4059   & 11.4  &  unblended \\
CH$_2$CCHCN & 36.8784700       & 7$_{1,6}-$6$_{1,5}$   & -5.3546  & 8.2  &  blended with U \\
CH$_2$CCHCN & 40.4257132     & 8$_{1,8}-$7$_{1,7}$    & -5.2292  & 9.9  & unblended  \\
CH$_2$CCHCN & 41.1870836     & 8$_{0,8}-$7$_{0,7}$   & -5.1981   & 8.9  & unblended  \\
CH$_2$CCHCN & 41.2976561     & 8$_{2,7}-$7$_{2,6}$   & -5.2225  &  13.4 & unblended  \\
CH$_2$CCHCN & 41.4207138     & 8$_{2,6}-$7$_{2,5}$   & -5.2187   &  13.4 & unblended  \\
CH$_2$CCHCN & 42.1384521     & 8$_{1,7}-$7$_{1,6}$   & -5.1751   & 10.2  &  blended with c-C$_3$H$_2$ \\
CH$_2$CCHCN & 45.4695206     & 9$_{1,9}-$8$_{1,8}$   & -5.0716  & 12.0  &  blended with CH$_3$NHCHO  \\
CH$_2$CCHCN & 46.2978836     & 9$_{0,9}-$8$_{0,8}$  & -5.0429  &  11.1 & unblended  \\
CH$_2$CCHCN & 46.4528399     & 9$_{2,8}-$8$_{2,7}$   & -5.0603   &  15.6 & blended with CH$_3$CONH$_2$  \\
CH$_2$CCHCN & 46.6280696     & 9$_{2,7}-$8$_{2,6}$   & -5.0555   & 15.7  & blended with
HC$_2$CHO  \\
CH$_2$CCHCN & 47.3948488     & 9$_{1,8}-$8$_{1,7}$   & -5.0176   & 12.5  & unblended  \\
CH$_2$CCHCN & 75.6463060      & 7$_{3,4}-$8$_{2,7}$    & -6.2551  & 17.0  & blended with U  \\
CH$_2$CCHCN & 75.6587572     & 15$_{1,15}-$14$_{1,14}$  & -4.3956  & 30.2  &  blended with U \\
CH$_2$CCHCN & 76.6665519     & 15$_{0,15}-$14$_{0,14}$   & -4.3767   & 29.6  & blended with C$_2$H$_5$OH  \\
CH$_2$CCHCN & 78.0996741     & 15$_{2,13}-$14$_{2,12}$   & -4.3600   & 34.4  &   blended with U \\ 
CH$_2$CCHCN & 78.8285795     & 15$_{1,14}-$14$_{1,13}$   & -4.3421  & 31.4  & blended with CH$_3$COOH  \\
CH$_2$CCHCN & 81.674936      & 16$_{0,16}-$15$_{0,15}$  & -4.2935   & 33.5  & blended with C$_2$H$_5$OH  \\
CH$_2$CCHCN & 84.0442448     & 16$_{1,15}-$15$_{1,14}$  & -4.2576  & 35.5  & unblended    \\
\hline
HCCCH$_2$CN  & 31.8489874    & 6$_{1,6}-$5$_{1,5}$  & -5.2773 &   6.2   & unblended   \\
HCCCH$_2$CN  & 33.8637212    & 6$_{1,5}-$5$_{1,4}$  & -5.1974 &  6.5    & unblended   \\ 
HCCCH$_2$CN  & 37.1392132    & 7$_{1,7}-$6$_{1,6}$  & -5.0691 &  8.0   & unblended  \\ 
HCCCH$_2$CN  & 38.1027037    & 7$_{0,7}-$6$_{0,6}$  & -5.0271 &   7.3   & unblended   \\ 
HCCCH$_2$CN  & 38.3423461    & 7$_{2,6}-$6$_{2,5}$ & -5.0555  &   10.6   & unblended  \\ 
HCCCH$_2$CN  & 38.6167091    & 7$_{2,5}-$6$_{2,4}$ & -5.0461  &  10.7    &  unblended \\ 
HCCCH$_2$CN  & 39.4865865    & 7$_{1,6}-$6$_{1,5}$  & -4.9892 &  8.4    & unblended  \\ 
HCCCH$_2$CN  & 42.0111351    & 5$_{1,5}-$4$_{0,4}$ & -5.4001  &  4.6    &  blended with n-C$_3$H$_7$CN \\ 
HCCCH$_2$CN  & 42.4217854    & 8$_{1,8}-$7$_{1,7}$  & -4.8900 &  10.0   &  unblended \\ 
HCCCH$_2$CN  & 43.8024266    & 8$_{2,7}-$7$_{2,6}$  & -4.8695 &   12.7   & blended with HNCO  \\ 
HCCCH$_2$CN  & 44.2102030    & 8$_{2,6}-$7$_{2,5}$  & -4.8574 &  12.8    & unblended  \\ 
HCCCH$_2$CN  & 45.0990808    & 8$_{1,7}-$7$_{1,6}$  & -4.8103 &  10.6    & unblended   \\ 
\hline
CH$_3$C$_3$N   &  33.0513004  & 8$_1-$7$_{1}$   & -5.3573   &  14.6 & \multirow{2}{*}{unblended} \\ 
CH$_3$C$_3$N   &  33.0516190  & 8$_0-$7$_{0}$  & -5.3505    & 7.1 &  \\ 
CH$_3$C$_3$N   &  37.1826557  & 9$_1-$8$_{1}$   & -5.1995   & 16.4 & \multirow{2}{*}{blended with CH$_3$OCHO} \\
CH$_3$C$_3$N   &  37.1830142  & 9$_0-$8$_{0}$   & -5.1942  & 8.9 & \\
CH$_3$C$_3$N   &  41.3139909  & 10$_1-$9$_{1}$  & -5.0590  & 18.4 & \multirow{2}{*}{blended with $aGg'$-(CH$_2$OH)$_2$}  \\
CH$_3$C$_3$N   &  41.3143891  & 10$_0-$9$_{0}$   &  -5.0546 & 10.9 & \\
CH$_3$C$_3$N   &  45.4453036  & 11$_1-$10$_{1}$  &  -4.9321  & 20.6 & \multirow{2}{*}{unblended}   \\
CH$_3$C$_3$N   &  45.4457416  & 11$_0-$10$_{0}$   &  -4.9286  & 13.1 &  \\
CH$_3$C$_3$N   &  49.5765916  &  12$_1-$11$_{1}$  & -4.8166  & 23.0 & \multirow{2}{*}{unblended}   \\
CH$_3$C$_3$N   &  49.5770694  &  12$_0-$11$_{0}$  & -4.8136  & 15.5 &  \\
CH$_3$C$_3$N   &  74.3636580   & 18$_1-$17$_{1}$  & -4.2809   & 41.4 & \multirow{2}{*}{unblended} \\
CH$_3$C$_3$N   &  74.3643850   & 18$_0-$17$_{0}$  & -4.2795  & 33.9 & \\
CH$_3$C$_3$N   &  82.6257600	 &  20$_1-$19$_{1}$  & -4.1422 & 49.1 & \multirow{2}{*}{blended with $s$-C$_2$H$_5$CHO}  \\
CH$_3$C$_3$N   &  82.6265180  &  20$_0-$19$_{0}$  & -4.1411 & 41.6 & \\
CH$_3$C$_3$N    & 86.756698   &  21$_1-$20$_{1}$  & -4.0780  & 53.3 & \multirow{2}{*}{blended with H$^{13}$CO$^+$}  \\
CH$_3$C$_3$N    & 86.757524  &  21$_0-$20$_{0}$   & -4.0770 & 45.8 &   \\
CH$_3$C$_3$N    & 90.8876208   &  22$_1-$21$_{1}$  & -4.0169   & 57.7& \multirow{2}{*}{blended with CH$_3$COCH$_3$}  \\
CH$_3$C$_3$N    & 90.8884956   &  22$_0-$21$_{0}$  & -4.0159   & 50.2 & \\
CH$_3$C$_3$N    & 95.0184892   &  23$_1-$22$_{1}$  & -3.9584   & 62.2& \multirow{2}{*}{unblended} \\
CH$_3$C$_3$N    & 95.0194037   &  23$_0-$22$_{0}$  & -3.9576   & 54.7 & \\
CH$_3$C$_3$N    & 99.1493060    &  24$_1-$23$_{1}$  & -3.9026  & 67.0 & \multirow{2}{*}{unblended} \\
CH$_3$C$_3$N    & 99.1502601   &  24$_0-$23$_{0}$   & -3.9017  & 59.5 & \\
CH$_3$C$_3$N    & 103.280069   &  25$_1-$24$_{1}$  & -3.8490   & 71.9 & \multirow{2}{*}{unblended}  \\
CH$_3$C$_3$N    & 103.2810626  &  25$_0-$24$_{0}$  & -3.8482   & 64.4 & \\
CH$_3$C$_3$N    & 107.4107758  & 26$_1-$25$_{1}$   & -3.7975   & 77.1 & \multirow{2}{*}{blended with CH$_3$CONH$_2$} \\
CH$_3$C$_3$N    & 107.4118089  & 26$_0-$25$_{0}$   & -3.7968   & 69.6 & \\
\hline
\label{tab:transitions_isomers}
\end{longtable}
{\small{$^{a}$  The format of the quantum numbers is $J_{K_a,K_c}$ for HCCCH$_2$CN and CH$_2$CCHCN (asymmetric rotors), and $J_{K}$ for CH$_3$C$_3$N (symmetric top molecule).}}

\vspace{3mm}

\vspace{3mm}
\subsubsection{Cyanopropyne (CH$_3$CCCN) }

Figure \ref{fig:molecules_ch3c3n} shows the spectra of multiple unblended or slightly blended transitions of CH$_3$CCCN (listed in Table \ref{tab:transitions_isomers}).
Unlike its isomers, which are asymmetric molecules, CH$_3$CCCN is a symmetric top molecule.
For the analysis, we have used the lowest energy $K$=0 and $K$=1 transitions (see Table \ref{tab:transitions_isomers}), which are the ones that dominate the line emission in a source with low $T_{\rm ex}$ like G+0.693 (5$-$20 K; see e.g. \citealt{zeng2018}). 
We fixed the FWHM  and $v_{\rm LSR}$ to the values derived for CH$_2$CCHCN, leaving $N$ and $T_{\rm ex}$ as free parameters.
We obtained a column density of (1.35$\pm$0.03)$\times$10$^{13}$ cm$^{-2}$ (Table \ref{tab:parameters}), and a molecular abundance of 1.0$\times$10$^{-10}$.
The isomeric ratios of CH$_2$CCHCN/CH$_3$CCCN and HCCCH$_2$CN/CH$_3$CCCN are $\sim$1.8 and $\sim$1.3, respectively.

\vspace{3mm}

\section{Discussion}
\label{sec:discussion}

\subsection{Interstellar chemistry}
\label{sec:discussion:chemistry}

\vspace{3mm}
\subsubsection{Oxygen-bearing nitriles}

\begin{figure}
\begin{center}
\includegraphics[width=17cm]{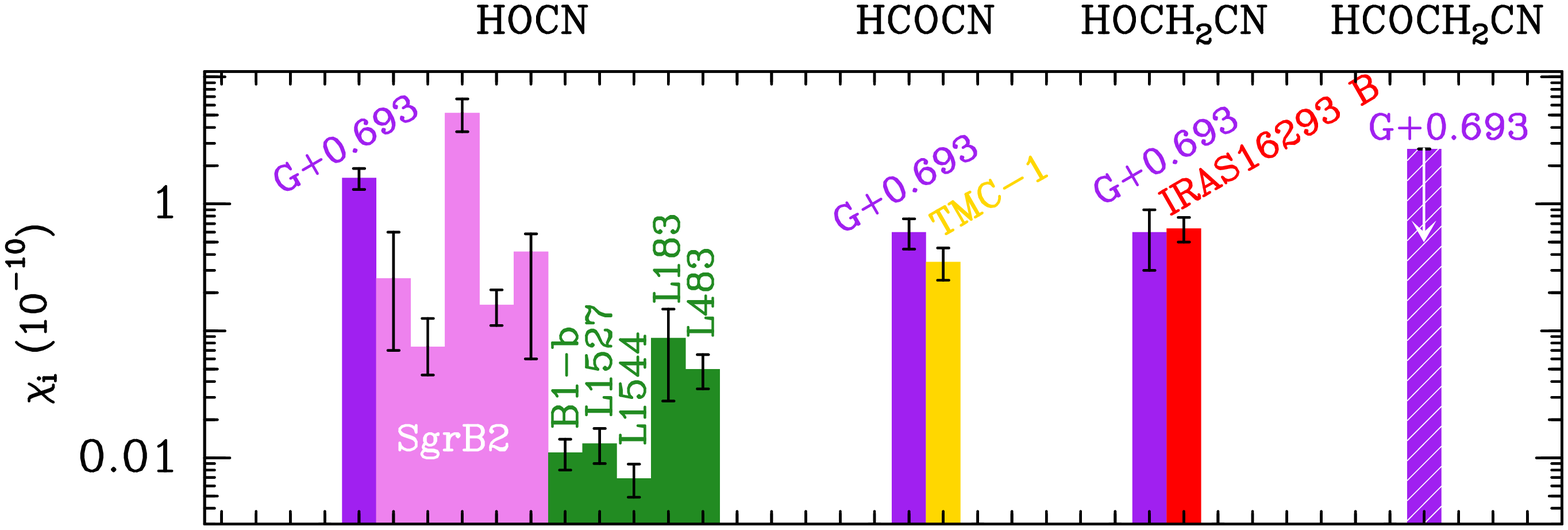}
\end{center}
\caption{Molecular abundances with respect to H$_2$ of the oxygen-bearing nitriles studied in this work derived in different interstellar sources. Purple bars correspond to G+0.693 (this work; see Table \ref{tab:parameters}), with the HCOCH$_2$CN value indicating an upper limit. We compare with other sources: several positions also in the Sgr B2 region (magenta; \citealt{brunken_interstellar_2010}, see also \citealt{marcelino_puzzling_2010}); several dense cores (B1-b, L1544, L183, L483) and the lukewarm corino L1527 (green: \citealt{marcelino_puzzling_2010,marcelino2018}); the dark cloud TMC-1 (yellow; \citealt{cernicharo2021_sulfur}); and the IRAS 16293-2422 B hot corino (red, \citealt{zeng2019}). To derive the uncertainties of the molecular abundances we have considered the uncertainties of the molecular column densities reported in the different works, or a 15$\%$ of the value of $N$ if the uncertainty is not provided, and we assumed an uncertainty for the $N$(H$_2$) column densities of 15$\%$.}
\label{fig:abundances-O-bearing}
\end{figure}

We show in Figure \ref{fig:abundances-O-bearing} the molecular abundances of the O-bearing nitriles detected towards G+0.693 studied in this work. The relative ratio of the detected species HOCN:HCOCN:HOCH$_2$CN is 2.8:1:1.   
By extrapolating the hydroxy/aldehyde (OH/HCO) ratio of HOCN/HCOCN to HOCH$_2$CN/HCOCH$_2$CN, one should expect an abundance of 0.15$\times$10$^{-10}$ for HCOCH$_2$CN, more than one order of magnitude lower than the upper limit derived from current observations ($<$2.7$\times$10$^{-10}$, see Table \ref{tab:parameters}). This suggests that deeper observations reaching higher sensitivity will be needed to address the detection of this species. 

In the following, we discuss possible formation routes of the different O-bearing nitriles, combining the results obtained in G+0.693 and in other interstellar sources with theoretical and experimental works:

{\bf $\bullet$ HOCN}: besides G+0.693, this species was detected previously towards several other positions of the Sgr B2 region in the Galactic Center (\citealt{brunken2009,brunken_interstellar_2010}),
and towards several dense cores (B1-b, L1544, L183, L483) as well as the lukewarm corino L1527 (\citealt{marcelino_puzzling_2010,marcelino2018}).
Figure \ref{fig:abundances-O-bearing} shows that the HOCN abundance derived in G+0.693 is of the same order of magnitude of those detected in other Sgr B2 positions ($\sim$10$^{-11}-$10$^{-10}$; \citealt{brunken_interstellar_2010}), and higher than those derived in the dense cores and L1527 (\citealt{marcelino2018}). 
This suggests that the role of surface-chemistry and the presence of shocks enhance the HOCN abundance, similarly to its isomer HNCO (\citealt{hasegawa1993,garrod_complex_2008,martin_tracing_2008,rodriguez-fernandez2010,quenard2018a}). The chemistry of the molecular clouds of the Galactic Center, and that of G+0.693 in particular, is dominated by large-scale shocks  (\citealt{martin-pintado2001,martin_tracing_2008}), which are responsible for the sputtering of dust grains, releasing many molecules formed on the grain surfaces into the gas phase (see \citealt{caselli1997,jimenez-serra2008}). This can increase the abundance of the species by orders of magnitude.
Similarly to isomer HNCO, which is efficiently formed on grain surfaces by hydrogenation of accreted OCN (\citealt{hasegawa1993,garrod_complex_2008}), HOCN can also be formed on grain mantles if the oxygen atom is hydrogenated: 
\begin{align}
\ch{
OCN + H &-> HOCN,
 }
\end{align}
and then subsequently released by shocks (\citealt{brunken_interstellar_2010}).
An alternative surface route might be the reaction of two highly abundant species: 
\begin{align}
\ch{
CN  + OH &-> HOCN
 }
\label{reaction:no+}
\end{align}

\begin{table}
\centering
\tabcolsep 5pt
\caption{Derived physical parameters of the nitriles towards G+0.693 analysed in this work using MADCUBA, along with their associated uncertainties. The fixed parameters used in the fit are shown without associated uncertainties (see text).}
\begin{tabular}{ c c  c c c c c  }
\hline
 Molecule  &$N^a$   &  $T_{\rm ex}$ & v$_{\rm LSR}$ & FWHM  & Abundance$^b$   \\
 &  ($\times$10$^{13}$ cm$^{-2}$) & (K) & (km s$^{-1}$) & (km s$^{-1}$) & ($\times$10$^{-10}$)    \\
\hline
HOCN  & 2.13$\pm$0.04  & 7.4$\pm$0.2  & 68.0$\pm$0.2  & 19.2$\pm$0.3 & 1.6     \\
HCOCN  & 0.76$\pm$0.11  & 7.4  & 68  & 19.2 & 0.6     \\
HOCH$_2$CN  & 0.8$\pm$0.2  & 7.4  & 67  & 19.2 & 0.6  \\
HCOCH$_2$CN  & $<$3.6 & 7.4  & 67  & 19.2 & $<$2.7     \\
\hline
CH$_2$CCHCN  & 2.34$\pm$0.06  & 12.1$\pm$0.5  & 66.1$\pm$0.3  & 21.3$\pm$0.7 & 1.7     \\
HCCCH$_2$CN  & 1.77$\pm$0.08  & 12.1  & 67.0$\pm$0.6  & 21.3 & 1.3   \\
CH$_3$CCCN  & 1.35$\pm$0.03  & 18.6$\pm$1.0  &  68 & 21.3  & 1.0    \\
\hline 
\end{tabular}
\label{tab:parameters}
\vspace{0mm}
\small
{\\ (a) The uncertainties of the column densities are derived by the AUTOFIT algorithm implemented in MADCUBA (see \citealt{martin2019} for details, and do not contain calibrations errors, which are expected to be $\sim$10$\%$.
 (b) We adopted $N_{\rm H_2}$=1.35$\times$10$^{23}$ cm$^{-2}$, from \citet{martin_tracing_2008}.
}
\end{table}

{\bf $\bullet$ HCOCN:} this species has been detected previously in the massive hot core SgrB2 (N) (\citealt{remijan_detection_2008}), and in the dark cloud TMC-1 (\citealt{cernicharo2021_sulfur}).
The HCOCN abundances found in G+0.693 and TMC-1 are very similar, in the range of (3.5$-$5)$\times$10$^{-11}$, as shown in Figure \ref{fig:abundances-O-bearing}. These two regions have very different physical conditions, which imprint their chemistry. While in the case of the dark and cold TMC-1 cloud gas-phase chemistry is thought to be dominant, since thermal or shock-induced desorptions are highly unlikely, the chemistry of G+0.693 is strongly affected by shocks, and thus surface chemistry also plays an important role.
Therefore, the similar HCOCN abundances in G+0.693 and TMC-1 points towards a predominant gas-phase chemistry origin. 
Indeed, the quantum chemical calculations by \citet{tonolo2020} have shown that HCOCN species can be efficiently formed through the gas-phase reaction between formaldehyde (H$_2$CO) and the cyanide radical (CN), which are highly abundant species in the ISM, in which the CN radical attacks the unsaturated carbon of H$_2$CO and substitutes one of the H atoms: 

\vspace{-10mm}
\begin{align}
\ch{
H_2CO  +  CN &-> HCOCN +  H ,
 }
\end{align}
\vspace{0mm}
%
{\bf $\bullet$ HOCH$_2$CN:} 
this species was first detected in the ISM towards the hot corino IRAS 16293-2422 B (\citealt{zeng2019}), and more recently towards the SMM1 hot corino in Serpens (\citealt{ligterink2021}).
The abundance derived in G+0.693 is 4.3$\times$10$^{-10}$, very similar to that derived in the hot component of IRAS 16293-2422 B (Figure \ref{fig:abundances-O-bearing}). 
The chemical model by \citet{zeng2019} considered the surface formation route proposed by the laboratory experiments of \citet{danger2012,danger_hydroxyacetonitrile_2013}:
\begin{align}
\ch{
    H_2CO + HCN &-> HOCH_2CN
}
\end{align}
and ion-neutral destruction reactions with H$_3^+$, HCO$^+$, and H$_3$O$^+$, and concluded that more chemical pathways are needed to explain the abundance observed in the hot corino IRAS 16293-2422 B. More recently, the quantum chemical cluster calculations performed by \citet{woon2021} have proposed new surface reactions between C$^+$, which is distributed throughout the whole Galactic Center (\citealt{harris2021}), and two very abundant species, HCN and HNC (e.g. \citealt{colzi2022}), embedded in H$_2$O icy grain mantles.
The C$^+$ ion reacts with HCN and HNC forming intermediate species that attacks neighboring H$_2$O molecules of the ices, resulting into the radicals HOCHNC and HOCHCN. These species can be easily hydrogenated on the grain surfaces to form HOCH$_2$CN.
The inclusion of these alternative surface routes in the chemical models might help to explain the HOCH$_2$CN abundances detected in G+0.693 and hot corinos, where the molecules can be injected to the gas phase through shocks and thermal effects, respectively.

{\bf $\bullet$ HCOCH$_2$CN:} 
the theoretical calculations  performed by \citet{horn2008} proposed that this species might be formed from two abundant precursors in the ISM:
\begin{align}
\ch{ 
   HC_3N + H_2O  &->  HCOCH_2CN
}
\end{align}
However, while this reaction might occur in aqueous solution, its activation energy, 216 kJ mol$^{-1}$ (25980 K), is too high to occur in the ISM. 
Recently, \citet{alessandrini2021} have studied the reaction between oxirane (or ethylene oxide, c$-$C$_2$H$_4$O) $-$also detected towards G+0.693 (\citealt{requena-torres_largest_2008}) $-$ and the CN radical. Although the main pathway is the H abstraction from oxirane, forming the oxiranyl radical, the formation of HCOCH$_2$CN + H is also possible with a rate of $\sim$10$^{-12}$ cm$^3$ molec$^{-1}$ s$^{-1}$.
New theoretical and/or experimental works of this species are needed to determine if it can be efficiently formed in the ISM, opening the possibility for its interstellar detection.

\vspace{5mm}
\subsubsection{C$_4$H$_3$N isomers}

\begin{figure}
\begin{center}
\includegraphics[width=17cm]{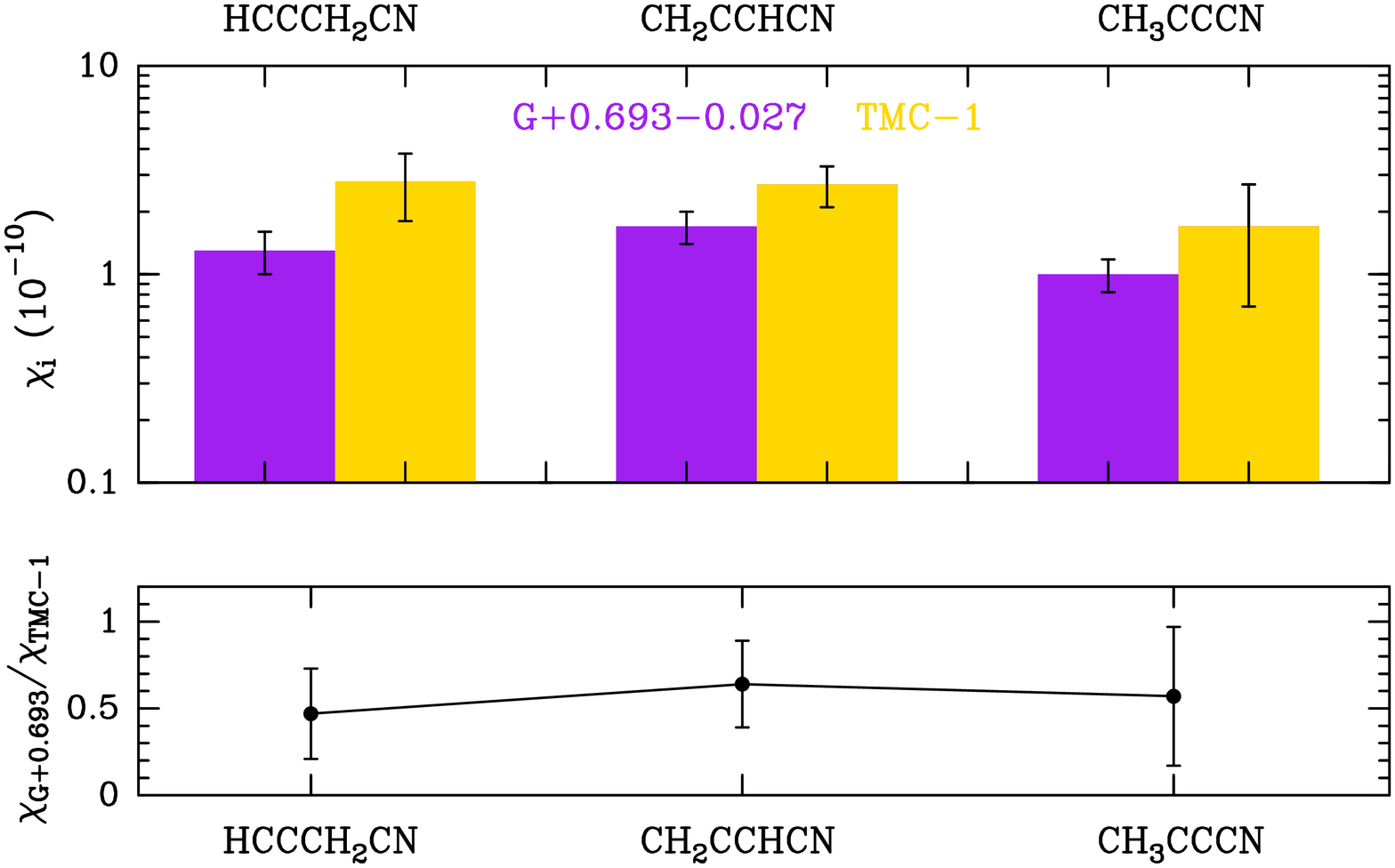}
\end{center}
\caption{{\it Upper panel:} Abundances with respect to H$_2$ of the C$_3$H$_4$N isomers detected towards G+0.693 (purple; this work) and TMC-1 (yellow, \citealt{marcelino2021}). To derive the uncertainties of the molecular abundances we have considered the uncertainties of the molecular column densities of the C$_3$H$_4$N isomers reported in this work (Table \ref{tab:parameters}) and in \citealt{marcelino2021}, and we assumed an uncertainty for the $N$(H$_2$) column density of 15$\%$. {\it Lower panel}: Molecular ratios between the abundances of the C$_3$H$_4$N isomers in G+0.693 and TMC-1.}
\label{fig:abundances-isomers}
\end{figure}

The unsaturated C$_4$H$_3$N isomers towards G+0.693 have very similar abundances within a factor of 2, spanning a range of (1.0$-$1.7)$\times$10$^{-10}$ (Table \ref{tab:parameters}), as also previously observed in the dark cloud TMC-1 by \citet{marcelino2021}. Moreover, Figure \ref{fig:abundances-isomers} shows that the abundances in these two molecular clouds, which have very different physical conditions, as mentioned above, are very similar.
This suggests that these molecules are predominantly formed through gas-phase chemistry (see previous discussion about HCOCN). Furthermore, since the three isomers are almost equally abundant, their respective formation might be linked to common precursors. Indeed, \citet{balucani2000} proposed that these unsaturated nitriles can be formed efficiently by reactions in which the cyanide radical (CN) attacks an unsaturated carbon of the hydrocarbons methylacetylene (CH$_3$CCH) and allene (CH$_2$CCH$_2$):
\begin{align}
 \ch{CH_3CCH + CN  &->[0.22/0.50] CH_3CCCN + H  \:, \\
     &->[0.0/0.50] CH_2CCHCN + H \:, }
\end{align}
\begin{align}
\ch{
CH_2CCH_2 + CN   &->[0.90] CH_2CCHCN + H \:, \\
               &->[0.10] HCCCH_2CN + H ;
}
\end{align}
%

in which the branching ratios for each reaction are indicated above each arrow (normalized to 1). These ratios were derived using the experiments and quantum chemical calculations by \citet{abeysekera2015} / \citet{balucani2000} in the first two reactions, and from \citet{balucani2002} in the latter two reactions.
These radical-neutral reactions show no entrance barriers, they have exit barriers well below the energy of the reactant molecules, and are exothermic.
The proposed precursors CN and CH$_3$CCH are abundant molecules in the ISM. In particular, they were detected towards G+0.693 with molecular abundances of 1.5$\times$10$^{-8}$ and 1.3$\times$10$^{-8}$, respectively (\citealt{rivilla2019a,bizzocchi2020}), so they are viable precursors. Allene (CH$_2$CCH$_2$) has zero dipole moment, so its detection through rotational spectroscopy is not possible, and thus its abundance is unknown. However, the similar abundances of the three isomers suggest that it can be as abundant as CH$_3$CCH in the ISM.

Regardless of the actual abundance of CH$_2$CCH$_2$, which is unknown, the proposed branching ratios seem to be in conflict with the observational findings in G+0.693 and TMC-1, since they are not able to produce equal abundance for the three isomers. As already noted by \citet{marcelino2021}, it would be interesting to study the branching ratios of the CH$_2$CHCH$_2$ + CN reaction using the chirped-pulse uniform flow experiment used by \citet{abeysekera2015} for the CH$_2$CCH + CN reaction, and compare them with the values derived from quantum chemical calculations by \citet{balucani2002}, to reconcile the experimental/theoretical works with the findings of the observations.

\vspace{5mm}
\section{Conclusions: Implications for the RNA-world}
\label{sec:discussion:role}

Compounds of the nitrile family, under early Earth conditions, offer a rich chemistry due to the large number of reactions that they can trigger.
Nitriles could be transformed into amides, carboxylic acids and esters via hydrolysis and alcoholysis respectively. Autocondensation of nitriles in a basic environment could yield to cyanoketones and cyanoenamines, a high reactive intermediate in the synthesis of complex five- and six-member heterocycles (\citealt{ayman1993}). The high amounts of ammonia of the reducing atmosphere of the primitive Earth is a favorable scenario to obtain amidines from nitriles (\citealt{shriner1944}). 
Moreover, the NCN backbone of amidines offer an unique structure to yield complex N-containing heterocycles like purine and pyrimidine nucleobases. Furthermore, nitriles can activate the formation of the building blocks of RNA, ribonucleotides (e.g. \citealt{powner2009,powner2010,patel2015}).
Two of the nitriles studied in this work, i.e. glycolonitrile and cyanoacetaldehyde, have been proposed as activation agents for the formation of more complex molecules with prebiotic relevance. The latter (HCOCH$_2$CN) is a precursor of cytosine (\citealt{robertson1995,nelson2001,menor2009}).
The former (HOCH$_2$CN) is not only a fundamental precursor to ribonucleotides and lipids (\citealt{ritson2012,ritson2013,patel2015,ritson2018,liu2018}), but also of other biologically-important molecules such as the simplest amino acid glycine (NH$_2$CH$_2$COOH; \citealt{rodriguez2019}), and of the nucleobase adenine through rapid HCN oligomerisation (\citealt{schwartz1982,menor2012}).
Unsaturated carbon-chain nitriles like the C$_4$H$_3$N isomers studied in this work are also especially interesting for prebiotic chemistry because the presence of unsaturated bonds allows further chemical evolution that can produce biomolecules (\citealt{rosi2018}).

This work extends the repertoire of nitriles detected in the G+0.693 molecular cloud, a region that exhibits one of the richest chemical content in the ISM, and hence it is a well suited testbed to census the molecular species present in the ISM. Besides HOCN, already reported by \citet{brunken_interstellar_2010} and \citet{zeng2018}, we have provided the tentative detections towards this source of HCOCN and HOCH$_2$CN (third detection in the ISM), and the detection of the three unsaturated C$_4$H$_3$N isomers (being the second source after TMC-1 in which all three isomers are identified). These detections confirm the rich reservoir of nitriles in space, and complete the list of prebiotic molecular precursors detected previously, including species directly involved in the synthesis of ribonucleotides such as glycolaldehyde (HCOCH$_2$OH; \citealt{hollis2004,requena-torres_organic_2006,beltran_first_2009,jorgensen2012}), urea (\citealt{belloche2019,jimenez-serra2020}), hydroxylamine NH$_2$OH (\citealt{rivilla2020b}), and 1,2-ethenediol (\citealt{rivilla2022a}); of amino acids, such as amino acetonitrile (NH$_2$CH$_2$CN; \citealt{belloche_detection_2008,melosso2020}); and of lipids, such as ethanolamine (NH$_2$CH$_2$CH$_2$OH; \citealt{rivilla2021a}), and propanol (CH$_3$CH$_2$CH$_2$OH; \citealt{jimenez-serra2022,belloche2022}).

In star- and planet-forming regions, this chemical feedstock can be processed through circumstellar disks, and subsequently incorporated into planetesimals and objects like comets and asteroids.  
We know that our planet suffered a heavy bombardment of extraterrestrial bodies $\sim$500 Myr after its formation (e.g. \citealt{marchi2014}).
Laboratory impact experiments have shown that a significant fraction of the molecules contained in comets and meteorites could have been delivered intact to the early Earth (\citealt{pierazzo1999,bertrand2009,mccaffrey2014,zellner2020,todd2020}). Once on the planetary surface, under the appropriate physical/chemical conditions, these molecules could have allowed the development of the prebiotic processes that led to the dawn of life on Earth.

\section*{Conflict of Interest Statement}

The authors declare that the research was conducted in the absence of any commercial or financial relationships that could be construed as a potential conflict of interest.

\section*{Author Contributions}

V.M.R initiated and led the project.
V.M.R, J.M.-P., F.R.-V., B.T. and P.d.V. performed the observations.
V.M.R., I.J.-S., and J.M.-P. performed the data reduction. 
V.M.R., L.C., S.Z., and I.J.S. contributed to the data analysis.
L.B. and M.M. performed the calculations of the cyanoacetaldehyde spectroscopy.
V.M.R. wrote an initial draft of the article.
All the authors, including J.G.dlC., S.M. and M.A.R.-T., participated in data interpretation and discussion.

\section*{Funding}
V.M.R. acknowledges support from the Comunidad de Madrid through the Atracción de Talento Investigador Modalidad 1 (Doctores con experiencia) Grant (COOL:Cosmic Origins of Life; 2019-T1/TIC-15379), and from the Agencia Estatal de Investigaci\'on (AEI) through the Ram\'on y Cajal programme (grant RYC2020-029387-I).
I.J.-S., J.M.-P and L.C. have received partial support from the Spanish State Research Agency (AEI) through project number PID2019-105552RB-C41.
J.G.d.l.C. acknowledges the Spanish State Research Agency (AEI) through project number MDM-2017-0737 Unidad de Excelencia “María de Maeztu”—Centro de Astrobiología and the Spanish State Research Agency (AEI) for partial financial support through Project No. PID2019-105552RB-C41. 

\section*{Acknowledgments}
We thank the two reviewers for providing very constructive and useful comments and suggestions, which contributed to improve our work. 
We also thank Dr. Rougal Ritson for interesting discussions about the relevance of nitriles in prebiotic chemistry.
We are very grateful to the IRAM 30m and Yebes 40m telescope staff for their precious help during the different observing runs. IRAM is supported by the National Institute for Universe Sciences and Astronomy/National Center for Scientific Research (France), Max Planck Society for the Advancement of Science (Germany), and the National Geographic Institute (IGN) (Spain). The 40m radio telescope at Yebes Observatory is operated by the IGN, Ministerio de Transportes, Movilidad y Agenda Urbana.
V.M.R. and L.C. have received funding from the Comunidad de Madrid through the Atracci\'on de Talento Investigador (Doctores con experiencia) Grant (COOL: Cosmic Origins Of Life; 2019-T1/TIC-15379). L.C. has also received partial support from the Spanish State Research Agency (AEI; project number PID2019-105552RB-C41).
P.d.V. and B.T. thank the support from the European Research Council (ERC Grant 610256: NANOCOSMOS) and from the Spanish Ministerio de Ciencia e Innovación (MICIU) through project PID2019-107115GB-C21. B.T. also acknowledges the Spanish MICIU for funding support from grant PID2019-106235GB-I00.

\section*{Data Availability Statement}

The data underlying this article will be shared on reasonable request to the corresponding author.

\bibliographystyle{frontiersinSCNS_ENG_HUMS} 
\bibliography{RNA}


\end{document}